\documentstyle[aps,pra,epsf]{revtex}
\begin{document}
\title{Classical 1D maps, quantum graphs and
       ensembles of unitary matrices}
\author{Prot Pako\'nski$^1$, Karol \.Zyczkowski$^2$ and Marek Ku\'s$^2$ \\
  \small $^1$Uniwersytet Jagiello\'nski,
	     Instytut Fizyki im. M.~Smoluchowskiego, \\
  \small ul. Reymonta 4, 30--059 Krak\'ow, Poland \\
  \small $^2$Centrum Fizyki Teoretycznej PAN, \\
  \small Al. Lotnik\'ow 32/44, 02--668 Warszawa, Poland \\
  \small e-mail addresses: \emph{pakonski@if.uj.edu.pl},
	 \emph{karol@cft.edu.pl}, \emph{marek@cft.edu.pl} }
\maketitle
\begin{abstract}
  We study a certain class of classical one dimensional piecewise
  linear maps. For these systems we introduce an infinite
  family of Markov partitions into equal cells. The symbolic
  dynamics generated by these systems is described by bistochastic
  (doubly stochastic) matrices. We analyze the structure
  of graphs generated from the corresponding symbolic dynamics.
  We demonstrate that the spectra of quantized graphs corresponding
  to the regular classical systems have locally Poissonian
  statistics, while quantized graphs derived from classically
  chaotic systems display statistical properties characteristic
  of Circular Unitary Ensemble, even though the corresponding
  unitary matrices are sparse.
\end{abstract}

\section{Introduction}
Although graphs served as models of physical systems for
a long time, the interest in properties of spectra of the
corresponding quantum systems has rather short history.
Quantum graphs were introduced as a ``toy'' model for studies
of quantum chaos by Kottos and Smilansky~\cite{Ko97,Ko99}.
They observed that spectral statistics of fully connected
graphs is well reproduced by random matrix theory (RMT) and
may be explained using the exact trace formula. Schanz and
Smilansky~\cite{Sc99} proved that the spectral correlation function
of simple graphs coincides with CUE expression for $2 \times 2$
matrices. The trace formula served to calculate the spectral
correlation function and to explain the deviations from
the RMT predictions for star graphs~\cite{Be99}. Tanner
introduced directed graphs and the corresponding unitary
transfer matrices~\cite{Ta00} and demonstrated fast
convergence of their spectral properties to RMT statistics
with increasing matrix size. A model for study of the
Anderson localization by means of quantum graphs was
introduced by Schanz and Smilansky~\cite{Sc00}, while
other properties of quantum graphs were recently analyzed
in~\cite{Ko00,Ba00}. Continuous dynamics on graphs was recently
discussed by Barra and Gaspard~\cite{Ba01}, who generated
Markov process by means of Poincar\'e surface of section.

In this work we propose another approach leading to quantum graphs.
We are interested in a certain class of one dimensional (1D)
piecewise linear maps with infinite family of Markov partition
into equal cells. For these systems we define a family of the
corresponding graphs by means of bistochastic (doubly stochastic)
transfer matrices $B$. If the transfer matrices are \emph{unistochastic}
(i.e. there exists a unitary matrix $U$ such that $B_{ij}=|U_{ij}|^2$)
the graphs may be quantized. We may thus establish a family of the
unitary transfer matrices by varying the lengths of bonds of graphs
and investigate their spectra. We demonstrate a link between the
character of dynamics of the classical system and the spectral
properties of the corresponding ensemble of unitary matrices.
For a class of regular 1D systems we show that the statistics
of the spectrum is locally Poissonian. If the classical system
has an arbitrary small, positive KS-entropy, the spectral statistics
of the corresponding ensemble of unitary matrices displays fluctuations
characteristic to Circular Unitary Ensemble (CUE)~\cite{Dy62,Me91}.
These matrices are sparse (only few nonzero elements in each row
and column) and by construction differ from typical random matrices
generated according to the Haar measure on $U(N)$.

The paper is organized as follows. In section~\ref{secsy}
we introduce the class of 1D maps of our interest.
Section~\ref{secrs} contains an analysis of the structure
and spectral properties of quantized graphs corresponding to
regular systems. In section~\ref{secsc} we study graphs
corresponding to Bernoulli shift and to some systems with
arbitrary small, but positive metric entropy. Section~\ref{secgs}
is devoted to study of generic chaotic systems, for which we find
CUE like spectral properties of the corresponding ensembles
of unitary matrices. The concluding remarks are presented in
section~\ref{secco}, while appendix~\ref{secum} contains the
condition necessary for a bistochastic matrix to be unistochastic
and the calculation of free parameters for averaging over
different graphs with the same topology. In appendix~\ref{secqm}
we show the construction of unitary matrices corresponding to
the classical map studied in section~\ref{secgs}.

\section{One dimensional dynamical systems} \label{secsy}
\begin{figure}[hbt]
  \hspace*{0.125\textwidth} \epsfxsize 0.65\textwidth \epsfbox{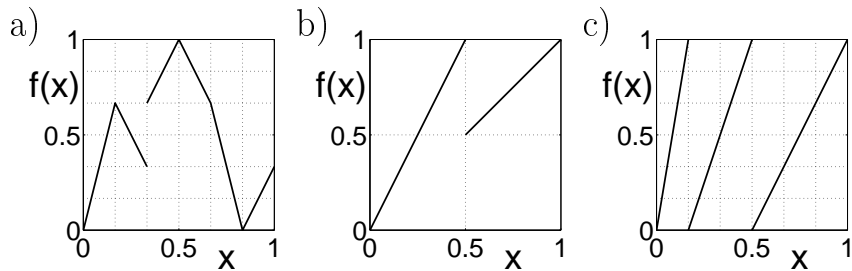}
  \caption{Examples of piecewise linear functions defining classical
    maps with Markov partition consisting of (a) $M=6$, (b) $M=2$
    and (c) $M=6$ equal cells. Only system (a) fulfils (i)--(iii).
    Observe that the 1D system shown in panel (b) is regular even
    though the average slope exceeds unity, this system does not
    fulfil the condition~(ii), the matrix $B$ corresponding to the
    system (c) is not unistochastic, so~(iii) is violated.}
  \label{figf}
\end{figure}

Consider a discrete one dimensional mapping acting on an interval
$I\subset I\!\!R$
\begin{equation}
  x_{n+1} = f(x_n), \label{1dmap}
\end{equation}
where $f:I\rightarrow I$ is piecewise linear. Assume that $f$ fulfils
three conditions:
\begin{enumerate}
\renewcommand{\labelenumi}{(\roman{enumi})}
\item there exists a Markov partition of the interval $I$ on $M$
  equal cells $E_i, i=1 \dots M$ and $f$ is linear on each cell $E_i$,
  \label{mcond}
\item for any $y \in I$ \label{bcond}
  \begin{equation}
    \sum_{x\in f^{-1}(\{y\})} \frac{1}{f'(x)} = 1 ,
  \end{equation}
\item the finite transfer matrix $B$, describing the evolution of measures
  uniform on the cells of Markov partition, under the action of
  the system $f$ (the Frobenius-Perron operator) is unistochastic.
  \label{ucond}
\end{enumerate}
Condition~(i) means that the graph of the function $f$
(later called \emph{diagram} of $f$ to avoid the confusion
with graphs corresponding to the map $f$) is composed of
straight lines (neither horizontal nor vertical) with
endpoints onto $M \times M$ grid on the square $I \times I$. Some
examples of functions fulfilling this condition are plotted in
figure~\ref{figf}. If we rescale $I$ to $[0,1]$ and write
$f(x)=c_jx+b_j$ for $x\in I_j$ where $\bigcup_j I_j=I$, then all $c_j$
have to be nonzero integers and also, the numbers $b_j$ and $|I_j|/|I|$
must be rational (multiplication by a sufficiently large $M$ makes all
these numbers integer). Any subpartition of the interval $I$ into $N=Mk$
equal cells ($k\in I\!\!N$) is also a Markov partition, so we
have an infinite family of Markov partitions for our models.
The Frobenius-Perron operator, reduced to densities (measures)
constant on each cell $E_j$, may be described by a matrix $B$
of finite size $M$. Let us write the elements of $B$ explicitly.
The system fulfilling the condition~(i) is fully characterized by
a set of $M$ linear functions
\begin{equation}
  f_i(x) = c_ix+b_i, \quad x \in \left[\frac{i-1}{M},\frac{i}{M}\right).
\end{equation}
By construction $c_i$ and $Mb_i$ are both integers. It is convenient to
express $c_i$ and $b_i$ in terms of two integers $l_i$ and $m_i$ as
\begin{equation}
  c_i = m_i-l_i, \quad b_i = \frac{m_i-i(m_i-l_i)}{M}.
\end{equation}
It is now easily seen that the only non-zero elements of $B$ in the
$i$-th row are
\begin{equation}
  B_{ij} = \frac{1}{|m_i-l_i|}, \quad \min(m_i,l_i)+1 \leq j
    \leq \max(m_i,l_i). \label{Belem}
\end{equation}
Closer inspection of the structure of the above formula reveals
the connection between the structure (positions of the non-zero
elements) of the matrix $B$ and the diagram of the system $f$
rotated clockwise by $\pi/2$.

The element $B_{ij}$ is equal to the probability of going from the cell
$E_i$ to $E_j$. From~(\ref{Belem}) $B$ is stochastic
($\sum_{j=1}^{N} B_{ij}=1$), all probabilities of going out
of the cell $E_i$ sum to unity. The probability of coming to the cell
$E_j$ from $E_i$ is equal to mean $1/|f'(x)|$ over $x\in E_i$ and
$f(x)\in E_j$. Thus due to condition~(ii) the probabilities of
transitions to any fixed cell $E_j$ also sum to one
($\sum_{i=1}^{N} B_{ij}=1$), so the transition matrix $B$ is
bistochastic. It is well known that all eigenvalues of a
bistochastic matrix are located in the unit circle, and the
leading eigenvalue is equal to one~\cite{Ma79}. The condition~(iii)
will be used in section~\ref{secgs}.

We may compute the KS-entropy of Markov chain generated by transition
matrix $B$~\cite{Ka95}
\begin{equation}
  H_{KS}(B)=-\sum_{i=1}^{N} \tilde p_i \sum_{j=1}^{N} B_{ij} \log B_{ij},
    \label{hks}
\end{equation}
where $\tilde p$ is the normalized left eigenvector of $B$
corresponding to the leading eigenvalue, $\tilde p B=\tilde p$,
$\sum_{i=1}^{N}\tilde p_i=1$. In condition~(i) we require that
$f$ is linear (not only piecewise linear) on each cell $E_j$. Then the
Markov partition on $M$ equal cells is a generating partition of the
system, so Eq.~(\ref{hks}) gives the dynamical entropy of the system.
The transition matrix is bistochastic in our case, so all components
of $\tilde p$ are nonzero ($\tilde p_i=1/N$) and $H_{KS}=0$ if and
only if all $B_{ij}\in\{0,1\}$. It is easy to see that if $\forall
x\in I: |f'(x)|=1$ the system is regular. We have showed that this
is also a necessary condition, provided the conditions~(i)
and~(ii) are fulfilled, since only the systems with $|f'|=1$
have all $B_{ij}\in\{0,1\}$. Note that in general there exist
regular systems with slopes larger than one, but they do not
satisfy the condition~(ii). An example of such a system is
plotted in Fig.~\ref{figf}(b).

\section{Regular systems} \label{secrs}
Partition of the interval $I$ determines the transition matrix
$B_{ij}$ describing the symbolic dynamics of the 1D map $f(x)$.
We introduce a directed graph with vertices corresponding to
each cell $E_i$ of the partition and bonds corresponding to all
nonzero elements of the transition matrix $B_{ij}$. Changing
the number $N$ of the cells in the partition we change the dimension
of the matrix $B^{(N)}$, so we may obtain a family of graphs
corresponding to any 1D map fulfilling the condition~(i).

\begin{figure}[hbt]
  \hspace*{0.15\textwidth} \epsfxsize 0.6\textwidth \epsfbox{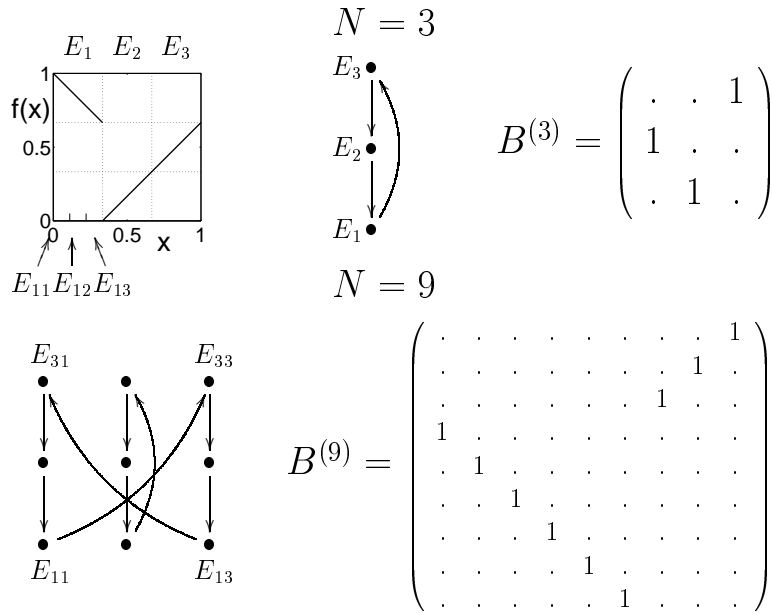}
  \caption{An example of a regular 1D system defined by function
    $f$ plotted in left top diagram, the corresponding graphs and
    the transition matrices representing Frobenius-Perron operators
    for Markov partitions into $N=3$ and $N=9$ cells. The dots
    represent zero elements of matrices. Note a link between the
    position of the nonzero elements of the matrices $B^{(N)}$ and
    the shape of the diagram of $f$ (rotated clockwise by $\pi/2$).}
  \label{figr}
\end{figure}
If we consider the regular systems with $|f'|=1$ the only nonzero
elements of matrices $B_{ij}$ are equal to 1. The matrices of this
type contain only one nonzero element equal to 1 in each row and
column and act as a permutation on a vector with $N$ elements. The
corresponding graph has only one incoming and one outgoing bond at
each vertex, so its valency is equal to one. It is composed from the
closed loops. We say that there is an inversion at vertex $j$ of the
graph if $f'=-1$ on the corresponding cell $E_j$ of the partition.
Dividing the interval $I$ into $M$ equal cells we find that the
longest loop of the graph has at most $M$ vertices. If we consider
a subpartition of $I$ into $N=Mk$ cells, each vertex of the graph
will be replaced by $k$ vertices of a new graph. The loops with an
even number of inversions will form $k$ new disconnected loops of the
same period. The loops with an odd number of inversions will create
$[k/2]$ disconnected loops with period doubled ($[\cdot]$ denotes
the integer part of a number) and also, if $k$ is odd, one loop of
the initial period. Fig.~\ref{figr} shows an example of a regular
system, for which the minimal Markov partition contains $M=3$ cells:
$E_1, E_2, E_3$. Due to $f'=-1$ in $E_1$ there exists one inversion
in the 3-vertices graph corresponding to this system. For larger $N$
the graph is composed of $[k/2]$ loops with 6-vertices and one
3-vertices loop for an odd $k$. The classical map has one periodic
orbit with period 3 and a continuous infinite family of orbits with
period 6. Note similarity in the spectrum of periods of periodic
orbits of the map and the graph for large odd $k$.

The quantization of a directed graph may be related to problem
of finding a unitary matrix $U$, such as $|U_{ij}|^2=B_{ij}$
for all indices $i,j$~\cite{Ta00}. The matrix $U$ may be seen
as a Floquet operator governing the discrete time evolution,
while $U^n$ represents the evolution operator at discrete time
$t=n$. Analyzing quantum maps, corresponding to time dependent
periodical systems, one studies the statistical properties of
phases of the unimodular eigenvalues of the unitary matrix
$U$~\cite{Ha91}. The elements of $U$ may be expressed by
$U_{ij}=r_{ij}e^{iL_{ij}}$, where the modulus are fixed by
the transition matrix $r_{ij}=\sqrt{B_{ij}}$, and the phases
$L_{ij}$, corresponding to lengths of the bonds of the graph,
are not specified by the corresponding 1D map. We only impose
that they fulfil the condition of unitarity $U^\dagger U =
{\mathbf{1}}$. Varying the phases $L_{ij}$ we obtain a family of
graphs -- an ensemble of unitary matrices. This corresponds to
varying the lengths of all bonds of the graph with its topology
preserved. We are interested in spectral properties of the
ensemble of unitary matrices defined in this way. If $B$ is a
permutation matrix, there is only one nonzero element in each
row and each column of $U$, so all phases $L_{ij}$ appearing
in matrix $U$ are uncorrelated.

The $N \times N$ unitary matrix $U$ corresponding to a regular
1D system is composed of $M$ square diagonal or antidiagonal
blocks of size $k$. For the system presented in Fig.~\ref{figr}
the matrix $U$ has the form
\begin{equation}
  U = \left(\begin{array}{ccc} 0 & 0 & V_1 \\
    V_2 & 0 & 0 \\ 0 & V_3 & 0 \end{array}\right) ,
\end{equation}
where $M=3$, $(V_1)_{ml}=\delta_{m,k+1-l}e^{iL_m}$ is an antidiagonal
$k \times k$ matrix, $(V_2)_{ml}=\delta_{ml}e^{iL_{m+k}}$ and
$(V_3)_{ml}=\delta_{ml}e^{iL_{m+2k}}$ are diagonal matrices. There are
$N$ free phases $L_m$ (the column index is superfluous here so we omit
it). The eigenvalues of block matrix $U$ may be expressed by means of
the spectra of its blocks $V_i$~\cite{Ga59}
\begin{equation}
  \mbox{eig}(U) = {}^M\sqrt{\mbox{eig}\left(V\right)}
    = {}^M\sqrt{\mbox{eig}\left(\prod_{i=1}^M V_i\right)} \ , \label{bleig}
\end{equation}
where the $M$-th root is a $M$ valued function. Unitary matrix
$V=\prod_{i=1}^M V_i$ of size $k$ is diagonal or antidiagonal,
depending on parity of the number of inversions in the corresponding
loop in the basic $M$-vertex graph. The nonzero elements of $V$
are equal to $e^{i{\mathcal{L}}_j}$, where ${\mathcal{L}}_j$,
$j=1 \ldots k$ are sums of different phases $L_m$, so
${\mathcal{L}}_j$ are uncorrelated. If $V$ is diagonal the spectrum
of $U$ consists of ${\{}e^{i({\mathcal{L}}_j+2\pi m)/M}{\}
}^{j=1 \ldots k}_{m=1 \ldots M}$. The eigenphases of $U$ belonging
to the same interval of the length $2\pi/M$ stem from different
series ${\mathcal{L}}_j$, so they are uncorrelated and the spectrum
has locally Poissonian statistics. If $V$ is antidiagonal, its
spectrum contains $\sqrt{e^{i({\mathcal{L}}_j+{\mathcal{L}
}_{k+1-j})}}$ and the spectrum of $U$ is composed of roots of
unimodular complex numbers of order $2M$. The character of the
spectrum remains Poissonian on the intervals of the length $\pi/M$.

\begin{figure}[hbt]
  \hspace*{0.075\textwidth} \epsfxsize 0.75\textwidth \epsfbox{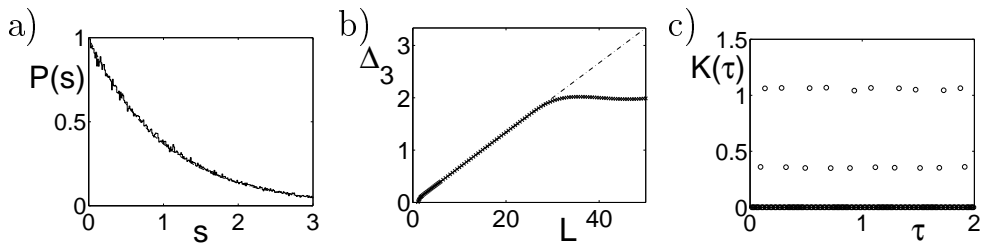}
  \caption{Spectral statistics for regular systems: (a) level spacing
    distribution $P(s)$, (b) spectral ridigity $\Delta_3(L)$ and (c)
    spectral form factor $K(\tau)$. The average is taken over an
    ensemble of $10^5$ matrices of the size $N=150$ with phases (bond
    lengths) chosen randomly. The statistics reveal deviations from
    the predictions of the Poissonian ensemble (dot dashed line) due
    to correlations on larger length scales.}
  \label{plotr}
\end{figure}
The above argument may be generalized for the situation, in which
the permutation matrix $B$ is reducible, composed of several
permutations with different periods $M_1 \ldots M_p$. In this case
the eigenphases of $U$ are uncorrelated in intervals of length
$\pi/\max M_i$. Fig.~\ref{plotr} shows the commonly used spectral
statistics~\cite{Me91,Ha91} calculated for the system shown in
Fig.~\ref{figr}. Note that the spectral ridigity $\Delta_3(L)$,
which measures correlations in the spectrum at the length scale
$L$, saturates at $L \approx 25 = N/6$ (there are $N=150$
eigenphases rescaled to the mean density constant and equal to $1$).
The divisor $6$ is equal twice the number of cells in minimal
Markov partition, since an inversion in one of the cells occurs
($f'(x)=-1$ for $x \in E_1$).

\section{Simple chaotic systems} \label{secsc}
\begin{figure}[hbt]
  \hspace*{0.15\textwidth} \epsfxsize 0.6\textwidth \epsfbox{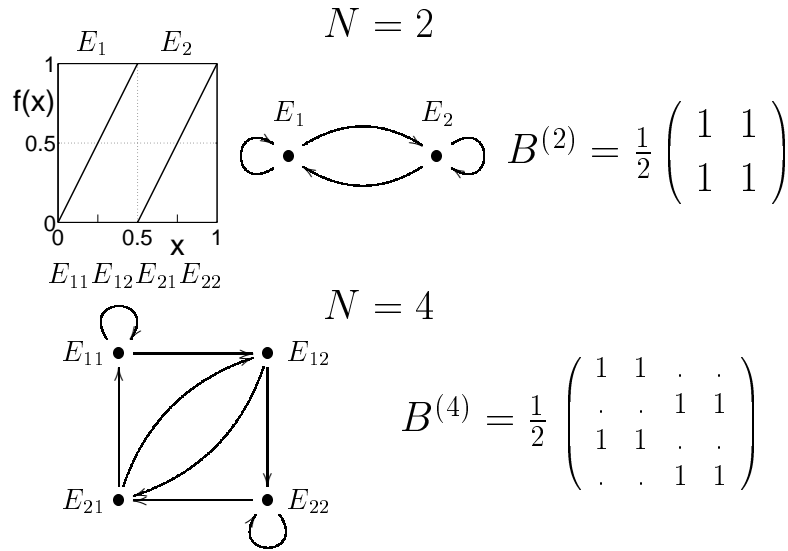}
  \caption{Bernoulli shift, the corresponding binary graphs and
    the matrices representing Frobenius-Perron operators for
    Markov partitions into $N=2$ and $N=4$ cells. Dots represent
    matrix elements equal to zero. For larger $N$ the structure
    of $B^{(N)}$ reflects the shape of $f(x)$.}
  \label{figt}
\end{figure}

The Bernoulli shift, $f(x)=2x \ \mbox{mod}\ 1$, is one of the simplest
1D maps displaying chaotic dynamics. This system fulfils the
conditions~(i), (ii) and~(iii). The graphs corresponding
to this system were analyzed by Tanner~\cite{Ta00}. The system,
graphs and matrices are plotted in Fig.~\ref{figt}. Existence of
a certain interval for which $|f'(x)|>0$ causes the mean valency
of the graphs to become larger than 1. Some vertices of the graphs
have more than one outgoing bond, and this fact is related with a
positive topological entropy of the corresponding classical map.
The number of such vertices grows proportionally with $N$ and
their presence is reflected in statistical properties of spectra
of the corresponding unitary matrices. For the binary graphs all
vertices have two outgoing bonds and as shown by Tanner~\cite{Ta00}
the spectral statistics converges fast to the predictions of CUE.
The traces of the unitary evolution operator $U^n$ may be
expressed by a sum over periodic orbits of the corresponding
classical graph, analogous to the semiclassical Gutzwiller trace
formula~\cite{Gu90}. For quantum graphs the formula
\begin{equation}
  \mbox{Tr}\ U^n = \sum_{\nu\in{\mathcal{PO}}(n)} A_\nu e^{iL_\nu} \ ,
    \label{posum}
\end{equation}
is an identity, where ${\mathcal{PO}}(n)$ is the set of periodic
orbits of period $n$, the amplitude $A_\nu=\prod_{(ij)\in\nu}\sqrt{B_{ij}}$
and the length $L_\nu$ is the sum of lengths of bonds belonging to the
periodic orbit $\nu$.

We will show that every periodic orbit of the graph corresponds
one to one to a periodic orbit of the dynamical system. Assume
that $\nu=j_1j_2\ldots j_n$ is a periodic orbit of period $n$
of the graph, where $j_i$ denote the vertices belonging to
$\nu$ and $|f'|>1$ on at least one of the cells $E_{j_c}$.
Let the functions
\begin{equation}
  F_{j_i}: E_{j_{i+1}} \ni x \rightarrow f^{-1}(x) \in E_{j_i}
\end{equation}
be the local inverse of $f$, where we identify $j_1$ and $j_{n+1}$.
The function $F_{j_c}$ is a contracting mapping since
$|F'_{j_c}|\le\frac{1}{2}$. We define
\begin{equation}
  F = F_{j_1} \circ F_{j_2} \circ \ldots \circ F_{j_n}
\end{equation}
which is a contraction of the cell $E_{j_1}$ onto itself, because
$|F'_{j_i}| \le 1$ for $i\neq c$. The function $F$ corresponds to
moving along the orbit backwards. Due to the Banach theorem there exists
exactly one fixed point of $F$ in $E_{j_1}$, namely $x_\nu=F(x_\nu)$.
Hence there is a periodic orbit of period $n$ in the dynamical system
corresponding to the orbit $\nu$ on the graph. The inverse is
also true, so there is one to one correspondence between the
orbits of the 1D chaotic system and the corresponding graphs.

\begin{figure}[hbt]
  \hspace*{0.075\textwidth} \epsfxsize 0.75\textwidth \epsfbox{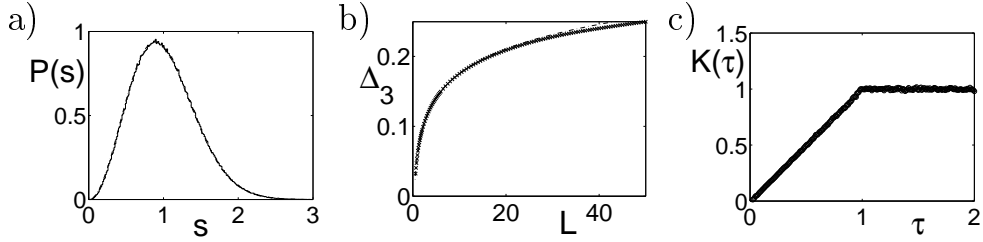}
  \caption{Spectral statistics for Tanner binary graphs averaged
    over bond lengths: (a) level spacing distribution $P(s)$, (b)
    spectral ridigity $\Delta_3(L)$ and (c) spectral form factor
    $K(\tau)$. The CUE prediction are represented by dot dashed lines
    (which almost coincide with the numerical data).}
  \label{plott}
\end{figure}

The sum~(\ref{posum}) goes over periodic orbits of the
classical chaotic map. The amplitudes $A_\nu$, built from
probabilities $B_{ij}$, are equal to $\prod_{i=0}^{n-1}
1/|f'(f^i(x_\nu))|^{1/2}$, so they are equal to the inverse
square roots of the instabilities of the orbits of the system.
Only the lengths $L_\nu$ are not determined yet, so we can
average over the lengths of the bonds imposing the constraints
due to unitarity of quantum propagator $U$. With growing
dimension $N$ the spectral statistics of quantized binary
graphs tends rapidly to the CUE statistics. Tanner observed
it analyzing spectral form factors for graphs with rationally
independent bond lengths~\cite{Ta00}. Fig.~\ref{plott} shows
the spectral statistics averaged over an ensemble of $10^5$
matrices of $N=100$, corresponding to graphs with randomly
chosen bond lengths (see appendix~\ref{secum}).

\begin{figure}[hbt]
  \hspace*{0.15\textwidth} \epsfxsize 0.6\textwidth \epsfbox{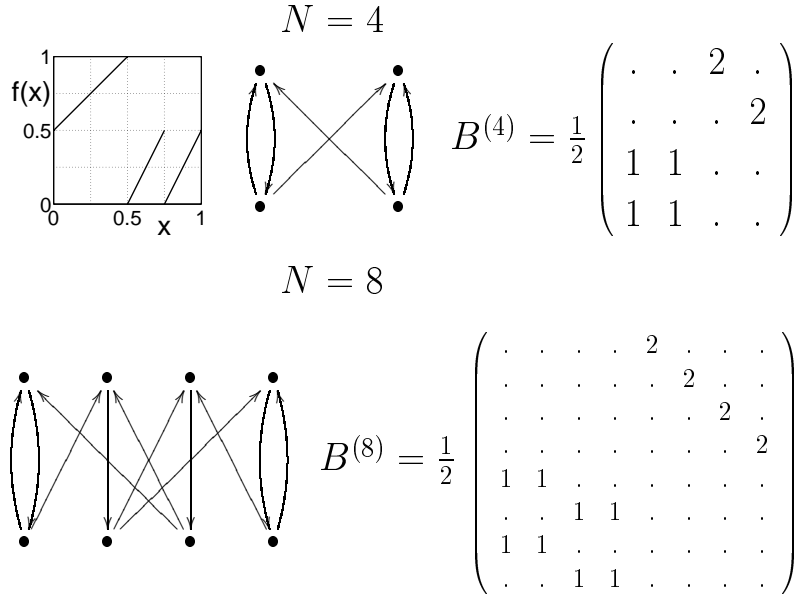}
  \caption{Chaotic system with $H_{KS}=\frac{1}{2}\ln 2$ and
    $\tilde M=2$, the function $f$ defining the system, the
    corresponding graphs and the transition matrices plotted
    for Markov partition into $N=4$ and $N=8$ cells. Zero
    elements of the matrices are marked by dots.}
  \label{figt2}
\end{figure}
\begin{figure}[hbt]
  \hspace*{0.075\textwidth} \epsfxsize 0.75\textwidth \epsfbox{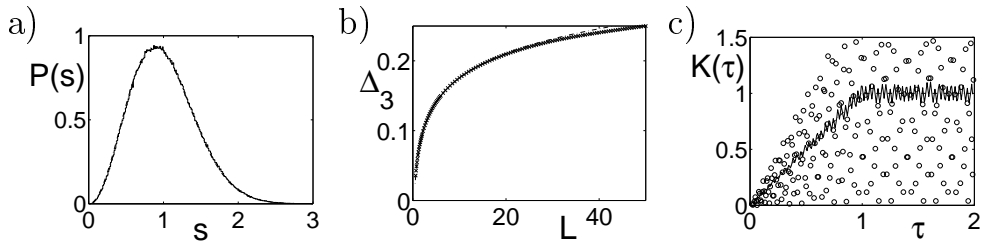}
  \caption{Spectral statistics for quantum graphs corresponding
    to the system shown in Fig.~\ref{figt2}: (a) level spacing
    distribution $P(s)$, (b) spectral rigidity $\Delta_3(L)$
    and (c) spectral form factor $K(\tau)$ averaged over $10^5$
    matrices of size $N=200$. Dot dashed lines represent CUE
    statistics. The form factor displays large fluctuations,
    although the average over a window of $\tau$ of size
    $\Delta\tau=0.07$ (solid line) is close to CUE results.}
  \label{plott2}
\end{figure}

We may also analyze a system, for which $|f'(x)|>1$ only
in certain regions $I_c \subset I$. The KS-entropy of such
a system is equal to the entropy of a system equivalent to
$f(x)$ in $I_c$ times the ratio $\mbox{vol}(I_c)/\mbox{vol}(I)$.
This allows us to construct a system with an arbitrary small,
positive KS-entropy. Consider the system presented in
Fig.~\ref{figt2} with dynamical entropy $H_{KS}=\frac{1}{2}\ln 2$
(equal to the topological entropy), as an illustration of such
situation. The structure of the unitary evolution operator $U$
is the same as the structure of the bistochastic transition matrix
$B$ which is determined by the shape of the diagram of $f$. Block
structure visible in Fig.~\ref{figt2} allows us to compute the
spectra as in Eq.~(\ref{bleig}), where $V$ is the product of all
blocks of $U$. If there is an even number of antidiagonal blocks,
the resulting matrix $V$ is a product of one matrix with Poisson
distribution of the spectrum and a certain number of non-diagonal
matrices with a CUE-like spectrum.
In ref.~\cite{Po98} we analyzed the spectra of composed ensembles
consisting of products of random matrices, each pertaining to a
given ensemble of unitary matrices. The CUE-like spectrum was found
to be robust with respect to the multiplication by matrices of other
ensembles. This is related to the fact that the Haar measure (CUE)
is invariant with respect to multiplication. Following these lines
of arguments, we conclude that the spectrum of $V$ has a CUE-like
properties. The spectrum of $U$ is equal to the $\tilde M$-th root
of the spectrum of $V$ ($\tilde M$ denotes the number of blocks in
$U$)
\begin{equation}
  e^{i\phi_U} = e^{i(\psi_j+2l\pi)/\tilde M}\ , \qquad
    j = 1\ldots k, \quad l = 0\ldots \tilde M-1,
\end{equation}
where $e^{i\psi_j}$ is the $j$-th eigenvalue of $V$, it consists
of $\tilde M$ copies of the spectrum of $V$ contracted $\tilde M$
times and placed one after another, so its statistical properties
are locally like these of CUE. Fig.~\ref{plott2} shows the
statistical properties of an exemplary chaotic system presented in
Fig.~\ref{figt2}, for which all spectra consist of $\tilde M=2$
identical parts of length $\pi$. Note that the level spacing
distribution and the spectral ridigity (for small $L$) show
good agreement with the predictions of RMT. On the other hand
the spectral form factor (i.e. the Fourier transform of the
two point correlation function of the spectral density~\cite{Ha91})
exhibits huge fluctuations since it measures correlations in
the entire spectrum and reveals the presence of 2 copies of
the same spectrum. However, the average taken over a small
interval of $\tau=n/N$ displays a CUE-like behaviour.

\begin{figure}[hbt]
  \hspace*{0.15\textwidth} \epsfxsize 0.6\textwidth \epsfbox{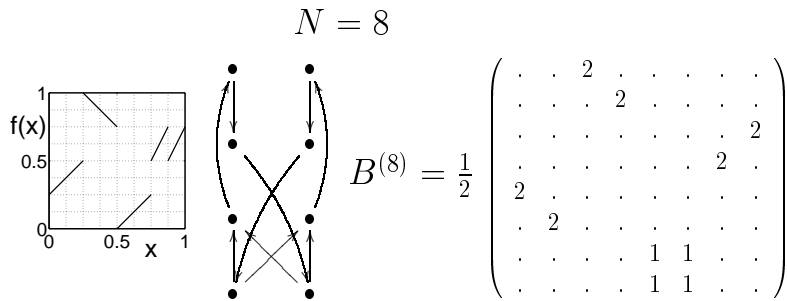}
  \caption{Chaotic system with $H_{KS}=\frac{1}{4}\ln 2$ and
    $\tilde M=4$, with an antidiagonal blocks in $B^{(N)}$:
    function $f$ defining the system, corresponding graph,
    and transition matrix. Find a link between its structure
    and the shape of $f(x)$.}
  \label{figt4}
\end{figure}

\begin{figure}[hbt]
  \hspace*{0.075\textwidth} \epsfxsize 0.75\textwidth \epsfbox{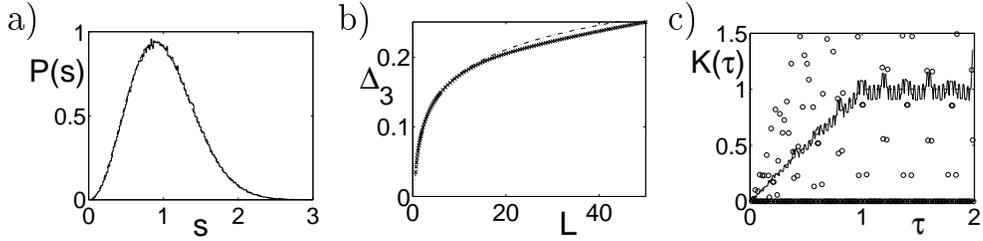}
  \caption{As in Fig.~\ref{plott2} for the system shown in
    Fig.~\ref{figt4}.}
  \label{plott4}
\end{figure}

Local agreement with the predictions of RMT is preserved even
though there exist an inversion in the map $f$, which implies
an antidiagonal blocks in $B^{(N)}$. To show this we analyzed
a system consisting of 4 blocks, presented in Fig.~\ref{figt4},
while the corresponding results are shown in Fig.~\ref{plott4}.

In the argument presented in this section we have assumed that
both parts of the system (one \emph{regular} with $|f'(x)|=1$
and the other \emph{chaotic} with $|f'(x)|>1$) are connected,
so the probability of going from one to the other is positive.
If this is not true, the system splits into two subsystems, one
regular and one chaotic. The spectrum of such a system will
contain a mixture of eigenvalues of both subsystems, so the
level spacing statistics could be described by an appropriately
adopted Berry-Robnik distribution~\cite{Be84}.

\section{Generic systems} \label{secgs}
A generic 1D piecewise linear system fulfilling the
conditions~(i), (ii) and~(iii) is chaotic and
$|f'(x)| \ge 2$ for all $x$ in $I$. The mean valency
of the corresponding graphs is also equal to or greater
than 2 and the topological entropy of these graphs
$H_{\mbox{top}}$ is not smaller than $\ln 2$. The
topological entropy of the map $f$ is equal to the log
of the spectral radius (absolute value of the largest
eigenvalue) of the \emph{connectivity matrix} $\tilde B$
\cite{Ka95}, where $\tilde B_{ij}=1$ if $B_{ij}>0$, and
$\tilde B_{ij}=0$ otherwise. It is worth to consider,
which of these graphs one may \emph{quantize} by
constructing corresponding family of unitary matrices.
It is equivalent to ask, which bistochastic transition
matrices $B$, are \emph{unistochastic}, so there exist
unitary $U$, such that $|U_{ij}|^2=B_{ij}$. It is clear
that not all dynamical systems fulfilling~(i) and~(ii)
give rise to bistochastic Frobenius-Perron matrices,
which are unistochastic. As an example let us consider
the system with $M=6$ shown in Fig.~\ref{figf}(c)
\begin{equation}
  \begin{array}{lr}
    f_1(x) = 6x, & \text{for } 0\leq x < \frac{1}{6}, \\ f_{2,3}(x)
    = 3x-\frac{1}{2}, & \text{for } \frac{1}{6}\leq x < \frac{1}{2}, \\
    f_{4,5,6}(x) = 2x-1, & \text{for } \frac{1}{2}\leq x \leq 1.
  \end{array}
\end{equation}
The corresponding transfer matrix
\begin{equation}
  B = \frac{1}{6} \left[ \begin{array}{cccccc}
    1 & 1 & 1 & 1 & 1 & 1 \\
    2 & 2 & 2 & 0 & 0 & 0 \\
    0 & 0 & 0 & 2 & 2 & 2 \\
    3 & 3 & 0 & 0 & 0 & 0 \\
    0 & 0 & 3 & 3 & 0 & 0 \\
    0 & 0 & 0 & 0 & 3 & 3 \end{array}\right]
\end{equation}
is clearly bistochastic but not unistochastic. Indeed one checks
that the necessary conditions for unistochasticity are not fulfilled
(see Appendix~\ref{secum} Eq.~(\ref{ucreq}), $k=3$ and $l=5$). This
is the smallest matrix corresponding to a dynamical system of the
considered type with this property, some other may be constructed for
$M=28=14+7+4+2+1$ or other perfect numbers. In appendix~\ref{secum}
we present a necessary condition for unistochasticity and discuss
some features of the ensemble of matrices $U$ corresponding to the
same bistochastic matrix $B$.

\begin{figure}[hbt]
  \hspace*{0.075\textwidth} \epsfxsize 0.75\textwidth \epsfbox{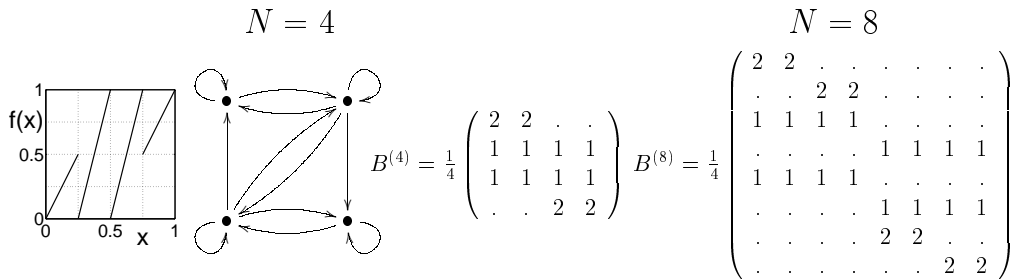}
  \caption{Classically chaotic 1D map (``four legs map''),
    corresponding graph and the transition matrices for Markov
    partitions into $N=4$ and $N=8$ cells. The slope of the map
    $f$ in each cell $E_i$ determines the size of elements in
    each row of the matrix $B^{(N)}$.}
  \label{figd}
\end{figure}
\begin{figure}[hbt]
  \hspace*{0.075\textwidth} \epsfxsize 0.75\textwidth \epsfbox{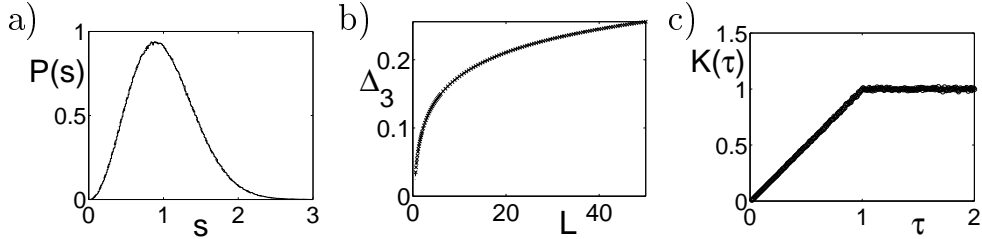}
  \caption{As in Fig.~\ref{plott2} for the four legs map shown in
    Fig.~\ref{figd}.}
  \label{plotd}
\end{figure}

One of the maps studied by Dellnitz et al.~\cite{De00} and
showed in Fig.~\ref{figd}, may serve as an example of a
classical map, for which transition matrices $B$ are
unistochastic (see Appendix~\ref{secqm}). Its
KS-entropy is equal to $\frac{3}{2}\ln 2$, while the topological
entropy equals to $\ln 3$. Instead of computing the spectral
radius of $\tilde B^{(4)}$ it is enough to observe that each
point in $I$ has exactly three preimages, and the topological entropy
may be computed as log of the mean number of preimages~\cite{Zy99}.
The average should be taken with respect to the Parry measure of
maximal entropy~\cite{Ka95}, which is not easy to specify in the
general case. However, since the number of preimages is constant,
its knowledge is not necessary in this case to obtain the required
result $H_{\mathrm{top}}=\ln 3$. Numerical investigations performed
for the corresponding ensemble of $10^5$ matrices of size $N=200$
confirmed a CUE-like character of spectral statistics in the entire
circle of eigenvalues as shown in Fig.~\ref{plotd}.

\begin{figure}[hbt]
  \hspace*{0.075\textwidth} \epsfxsize 0.75\textwidth \epsfbox{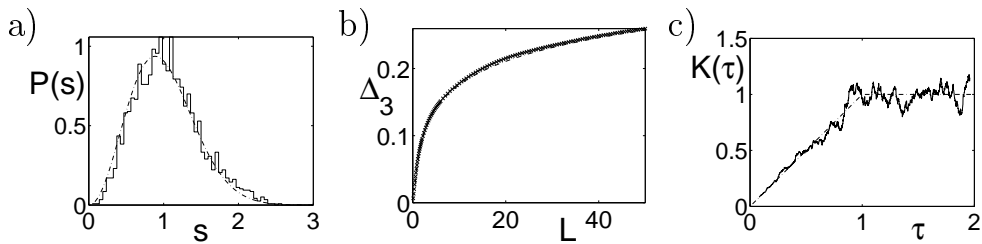}
  \caption{As in Fig.~\ref{plott2} for the four legs map shown in
    Fig.~\ref{figd}, without averaging over the ensemble of unitary
    matrices. The statistics calculated from single matrix of size 2500
    with randomly chosen free phases are plotted.}
  \label{plote}
\end{figure}

Instead of averaging over an entire ensemble of unitary matrices with
random phases, we may study spectral statistics for one concrete member
of the ensemble. Figure~\ref{plote} shows the spectral statistics for
a single matrix $U$, in which the phases, not determined by unitarity,
being set randomly (see Appendix~\ref{secum}). Observe an agreement
with CUE predictions (the form factor $K(\tau)$ is averaged over the
window of $\tau$, $\Delta\tau=0.08$). Interestingly, for this system
a CUE-like behaviour was also found for a single matrix $U_0$
(corresponding to the map plotted in Fig.~\ref{figd}) with all free
parameters set to zero (thus exhibiting strong correlations between
phases). Matrix $U_0$ is real (therefore orthogonal).

Although a single member of the ensemble displays CUE-like spectrum,
the averaging over free phases provides larger statistics. The form
factor may be expressed as a double sum over periodic orbits
\begin{equation}
  K(\tau) = \frac{1}{N}\left\langle |\mbox{Tr}\ U^n|^2 \right\rangle
    = \frac{1}{N} \left\langle \sum_{\nu,\nu'\in{\mathcal{PO}}(n)}
    A_\nu A_{\nu'} e^{i(L_\nu-L_{\nu'})} \right\rangle , \label{ffac}
\end{equation}
where $\tau=n/N$ and the averaging goes over all allowed
realizations of the graph (all bond lengths $L_{ij}$, for
which unitarity of $U$ is preserved). The difference
of the lengths of orbits is equal to $L_\nu-L_{\nu'} =
\sum_{(ij)\in\nu} L_{ij}-\sum_{(lm)\in\nu'} L_{lm}$.
Inserting new variables $\tilde L_{ij}=L_{ij}-L_{i1}-L_{1j}+L_{11}$
we have
\begin{equation}
  L_\nu-L_{\nu'} = \sum_{(ij)\in\nu} (\tilde L_{ij}+L_{i1}+L_{1j}-L_{11})
    - \sum_{(lm)\in\nu'} (\tilde L_{lm}+L_{l1}+L_{1m}-L_{11}),
\end{equation}
where the lengths $\tilde L_{ij}, \ i,j=2 \ldots N$ are fixed by
the condition assuring unitarity of $U$ (see Appendix~\ref{secum}),
while $(2N-1)$ remaining phases $L_{i1}$ and $L_{1j}$ are drawn
randomly with a uniform distribution on $[0,2\pi)$. Due to
averaging over $L_{i1}$, $L_{1j}$ the only non vanishing terms
in the sum~(\ref{ffac}) arise from periodic orbits $\nu$ and
$\nu'$ of the same \emph{degeneracy class} which visit each
vertex of the graph the same number of times. This observation
generalizes results of Tanner for binary graphs~\cite{Ta00}.
Therefore the calculation of $K(\tau)$ can be reduced to
a combinatorial problem after finding the fixed lengths
$\tilde L_{ij}$. The calculations of form factor for fully
connected graphs~\cite{Ko99} and binary graphs~\cite{Ta00}
shows fast convergence of the spectral statistics to RMT
predictions with increasing matrix size $N$.

\section{Conclusions} \label{secco}
We proposed to study a certain class of piecewise linear 1D
dynamical systems, for which the Markov transition matrix
(of size $M$) is bistochastic. It has a structure of the
diagram of $f$, defining the classical map, rotated clockwise by
$\pi/2$. For each subpartition of the Markov partition we
construct a graph and a family of unitary matrices, of size
equal to a multiplicity of $M$. This procedure might be thus
considered as a ``quantization'' of a $1D$ classical maps,
in a sense that it produces a family of corresponding unitary
matrices. The above quotation marks represent the fact that
no canonical action-angle quantization of classical systems
with 1D phase space is possible.

We demonstrated that the spectral properties of ensembles of
unitary matrices defined in this way reflect the character of
the classical dynamics. For regular maps the graphs are
composed of several finite disconnected loops and the spectral
statistics of the corresponding unitary evolution operators
is (locally) Poissonian. On the other hand, chaotic systems
(with arbitrarily small, but positive metric entropy), lead to
ensembles of unitary matrices with (locally) CUE-like spectral
fluctuations. These unitary matrices are sparse (only a few
non zero elements in each row and column), in contrast to the
random matrices, typical with respect to the Haar measure on
$U(N)$.

We have shown that the classical 1D system and the corresponding
family of graphs are topologically conjugated, since each
periodic orbit of the map corresponds to exactly one periodic
orbit of each graph. Applying finer and finer Markov partitions
to the system, introduces a natural limit in the family of the
corresponding graphs. It is the limit of number of vertices
tending to infinity with the set of periodic orbits preserved.
This seems to be the correct limit to reach random matrix theory
predictions for the spectral statistics of quantized graphs.
Spectral properties of a graph may be
expressed by the exact trace formula based on a sum over periodic
orbits. This sum goes over all periodic orbits of the 1D system
and the amplitudes are equal to inverse square roots of the
instabilities of each orbit. The phases in the sum remain free
parameters of the model, with constraints due to unitarity of
the evolution operator. We find conditions necessary for
a bistochastic transfer matrix to be unistochastic, which
allow us to perform averaging over random phases (bond
lengths) with unitarity preserved.

\section*{Acknowledgments}
Financial support by Polish KBN grant no 2~P03B~009~18 (P.P.) and
no 2~P03B~072~19 (K.\.Z. and M.K.) is gratefully acknowledged. PP
is grateful to H.~Schanz for an invitation and a financial support
which enabled him to attend the Conference on \emph{Random Network
Models} organized in G\"otingen in December 2000.

\section*{Note added in proof}
After this work was completed we learnt about very recent related
works on unistochastic matrices of Berkolaiko~\cite{Be01} and
Tanner~\cite{Ta01}.

\appendix
\section{Unistochastic matrices} \label{secum}
For a real bistochastic matrix $B$ of size $N$
\begin{equation}
  B_{jk} \ge 0, \quad \sum_{j=1}^N B_{jk}=1, \quad \sum_{k=1}^N B_{jk}=1,
    \label{beqs}
\end{equation}
we want to find a unitary matrix $U$, such that $|U_{jk}|^2=B_{jk}$.
We can manipulate with $N^2$ variable phases $L_{jk}$ of the matrix
$U$
\begin{equation}
  U_{jk} = \sqrt{B_{jk}} \,\, e^{iL_{jk}} \ .
\end{equation}
The unitarity of $U$ requires that
\begin{equation}
  U^\dagger U={\mathbf{1}} \quad\Leftrightarrow\quad \sum_{j=1}^N
    \sqrt{B_{kj}B_{lj}} \,\, e^{i(L_{lj}-L_{kj})}=\delta_{kl} \ .
    \label{ueqs}
\end{equation}
Equations~(\ref{ueqs}) for $k=l$ are fulfilled due to stochasticity
of $B$, Eq.~(\ref{beqs}). Equations~(\ref{ueqs}) for $k<l$ are
complex conjugated of these for $k>l$. Transforming the system of
equations~(\ref{ueqs})
\begin{equation}
  U(U^\dagger U-{\mathbf{1}})U^\dagger =
    UU^\dagger(UU^\dagger-{\mathbf{1}}) = 0 \quad
    \Rightarrow \quad UU^\dagger-{\mathbf{1}} = 0
  \label{veqs}
\end{equation}
we found that $(N-1)$ equations of the system are depending on
the other ones, because the diagonal equations of the righthand
side are equivalent to conditions for bistochasticity of $B$ --
the first sum in Eq.~(\ref{beqs}). We subtracted the unity from
$N$ since there is one common condition stemming from each of the
sums in Eq.~(\ref{beqs}), $\sum_{ij=1}^N B_{ij}=N$. The number of
independent real equations in~(\ref{ueqs}) equals thus $(N-1)^2$.

We introduce new variables $\tilde L_{kj}=L_{kj}-L_{k1}-L_{1j}+L_{11}$,
so the equations~(\ref{ueqs}) are equivalent to
\begin{equation}
  \sum_{j=1}^N \sqrt{B_{kj}B_{lj}} \,\,
    e^{i(\tilde L_{lj}-\tilde L_{kj})}=\delta_{kl} \ . \label{leqs}
\end{equation}
Since $\tilde L_{k1}=\tilde L_{1j}=0$, so the $(N-1)^2$
equations~(\ref{ueqs}) depend on $(N-1)^2$ real independent
variables $\tilde L_{kj},\ k,j = 2 \ldots N$. These equations
give the values of $\tilde L_{kj}$. For $N=2$ there exists
only one phase $\tilde L_{22}$ equal to $\pi$ as shown by
Tanner~\cite{Ta00}. This value does not depend on the
bistochastic matrix $B_{jk}$. Eqs.~(\ref{leqs}) do not depend
on $L_{k1}$ and $L_{1j}$, so we may average over them,
assuming the uniform distribution over the unit circle,
$P(L)=\frac{1}{2\pi}$ for $L \in [0,2\pi)$. This is
equivalent to averaging of the matrix $U=D_1 \tilde U D_2$
over diagonal unitary random matrices $D_1$ and $D_2$,
with a fixed matrix $\tilde U$, fulfilling
$|\tilde U_{jk}|^2=B_{jk}$.

The system of equations~(\ref{ueqs}) is solvable only if
\begin{equation}
  \forall\,k \neq l \quad \max_{m=1 \ldots N} \sqrt{B_{km}B_{lm}}
    \le \frac{1}{2} \sum_{j=1}^N \sqrt{B_{kj}B_{lj}} \quad
    \mbox{and} \quad \max_{m=1 \ldots N} \sqrt{B_{mk}B_{ml}}
    \le \frac{1}{2} \sum_{j=1}^N \sqrt{B_{jk}B_{jl}} \label{ucreq}
\end{equation}
which is the condition for the possibility of balance the maximal
term in the sums~(\ref{ueqs}) by the rest and the analog condition
for the sums stemming from~(\ref{veqs}). This necessary condition
for a bistochastic matrix to be unistochastic becomes sufficient
in $3 \times 3$ case. For $N=2$ all bistochastic matrices are
orthostochastic (and therefore also unistochastic). However for
$N \ge 3$ several examples of bistochastic matrices, which are not
unistochastic, are known~\cite{Ma79}, they do not satisfy the
condition~(\ref{ucreq}). Statistical properties of random
unistochastic matrices are studied in~\cite{Zy01}.

\section{Construction of a family of quantum maps} \label{secqm}
The construction of quantum graphs presented in section~\ref{secgs}
hinge on the possibility of finding an unitary (or orthogonal)
matrix corresponding to the bistochastic matrix obtained for a
given dynamical system for an arbitrary Markov subdivision of the
original partition. Rather then attempting the formulation of
the most general theorem concerning the existence of such
unistochastic matrices, we present here a construction for a 
particular case of the ``four legs map'' presented in Fig.~\ref{figd}.
In other words we show that the all bistochastic matrices $B^{(N)}$
corresponding to this classical dynamical system are unistochastic.

The original bistochastic matrix for the ``four legs map'' introduced
in~\cite{De00} for the initial Markov partition into four cells reads
\begin{equation}
  B^{(4)} = \frac{1}{4}\left( \begin{array}{cccc}
	2 & 2 & 0 & 0 \\
	1 & 1 & 1 & 1 \\
	1 & 1 & 1 & 1 \\
	0 & 0 & 2 & 2 \end{array} \right) \ .
\end{equation}

Let us consider the following choice of a corresponding unitary
matrix
\begin{equation}
  U^{(4)} = \frac{1}{2}\left( \begin{array}{cccc}
	\sqrt{2} & -\sqrt{2} & 0 & 0 \\
	1 & 1 & 1 & 1 \\
	1 & 1 & -1 & -1 \\
	0 & 0 & \sqrt{2} & -\sqrt{2} \end{array} \right) \ .
\end{equation}
The bistochastic matrix for the two times finer subdivision reads
\begin{equation}
  B^{(8)} = \frac{1}{4}\left( \begin{array}{cccccccc}
	2 & 2 & 0 & 0 & 0 & 0 & 0 & 0 \\
	0 & 0 & 2 & 2 & 0 & 0 & 0 & 0 \\
	1 & 1 & 1 & 1 & 0 & 0 & 0 & 0 \\
	0 & 0 & 0 & 0 & 1 & 1 & 1 & 1 \\
	1 & 1 & 1 & 1 & 0 & 0 & 0 & 0 \\
	0 & 0 & 0 & 0 & 1 & 1 & 1 & 1 \\
	0 & 0 & 0 & 0 & 2 & 2 & 0 & 0 \\
	0 & 0 & 0 & 0 & 0 & 0 & 2 & 2 \end{array} \right) \ .
\end{equation}
Using our knowledge of the unitarity of $U^{(4)}$ we want to
find a unitary matrix $U^{(8)}$, such that $B^{(8)}_{ij}=
|U^{(8)}_{ij}|^2$. This can be done by replicating twice
the rows of $U^{(4)}$ with an appropriate shift. Thus
$U^{(8)}$ may read as follows
\begin{equation}
  U^{(8)} = \frac{1}{2}\left( \begin{array}{cccccccc}
	\sqrt{2} & -\sqrt{2} & 0 & 0 & 0 & 0 & 0 & 0 \\
	0 & 0 & \sqrt{2} & -\sqrt{2} & 0 & 0 & 0 & 0 \\
	1 & 1 & 1 & 1 & 0 & 0 & 0 & 0 \\
	0 & 0 & 0 & 0 & 1 & 1 & 1 & 1 \\
	1 & 1 & -1 & -1 & 0 & 0 & 0 & 0 \\
	0 & 0 & 0 & 0 & 1 & 1 & -1 & -1 \\
	0 & 0 & 0 & 0 & \sqrt{2} & -\sqrt{2} & 0 & 0 \\
	0 & 0 & 0 & 0 & 0 & 0 & \sqrt{2} & -\sqrt{2}
	\end{array} \right) \ .
\end{equation}
Note that the number of the non zero elements in $U^{(8)}$
is two times larger than the number of the non zero elements
in $U^{(4)}$. By replicating each row of the matrix $U^{(4)}$
$k$ times we obtain in the similar way unitary matrices of the
size $4k$.

The success of this construction is due to the fact that the
nonzero elements of the first and fourth rows of $U^{(4)}$
are equal, which implies orthogonality of the following
pairs of rows in $U^{(8)}$: (2,3), (2,5), (4,7) and (6,7).
However this is not the only way of constructing the family
of unitary matrices $U^{(N)}$. For example we found another
family $\tilde U^{(N)}$ originating from
\begin{equation}
  \tilde U^{(4)} = \frac{1}{2}\left( \begin{array}{cccc}
	\sqrt{2} & -\sqrt{2} & 0 & 0 \\
	1 & 1 & -1 & 1 \\
	1 & 1 & 1 & -1 \\
	0 & 0 & \sqrt{2} & \sqrt{2} \end{array} \right) \ ,
\end{equation}
even though the procedure described above fails in this case.
This fact allows us to hope that the class of classical 1D
maps leading to quantum graphs (fulfilling the conditions (i),
(ii) and (iii)) is not trivial.

\end{document}